*Article*

# STANet: A Novel Spatio-Temporal Aggregation Network for Depression Classification with Small and Unbalanced FMRI Data

Wei Zhang [1], Weiming Zeng [1,*], Hongyu Chen [1], Jie Liu [1], Hongjie Yan [2], Kaile Zhang [3], Ran Tao [3], Wai Ting Siok [3] and Nizhuan Wang [3,*]

[1] Lab of Digital Image and Intelligent Computation, College of Information Engineering, Shanghai Maritime University, Shanghai 201306, China; 202230310117@stu.shmtu.edu.cn (W.Z.); 202330310129@stu.shmtu.edu.cn (H.C.); 202230310126@stu.shmtu.edu.cn (J.L.)
[2] Department of Neurology, Affiliated Lianyungang Hospital of Xuzhou Medical University, Lianyungang 222002, China; yanhjns@gmail.com
[3] Department of Chinese and Bilingual Studies, The Hong Kong Polytechnic University, Hong Kong, China; kaile-keller.zhang@polyu.edu.hk (K.Z.); ran.tao@polyu.edu.hk (R.T.); wai-ting.siok@polyu.edu.hk (W.T.S.)
* Correspondence: zengwm86@163.com (W.Z.); wangnizhuan1120@gmail.com (N.W.)

**Abstract: Background**: Early diagnosis of depression is crucial for effective treatment and suicide prevention. Traditional methods rely on self-report questionnaires and clinical assessments, lacking objective biomarkers. Combining functional magnetic resonance imaging (fMRI) with artificial intelligence can enhance depression diagnosis using neuroimaging indicators, but depression-specific fMRI datasets are often small and imbalanced, posing challenges for classification models. **New Method**: We propose the Spatio-Temporal Aggregation Network (STANet) for diagnosing depression by integrating convolutional neural networks (CNN) and recurrent neural networks (RNN) to capture both temporal and spatial features of brain activity. STANet comprises the following steps: (1) Aggregate spatio-temporal information via independent component analysis (ICA). (2) Utilize multi-scale deep convolution to capture detailed features. (3) Balance data using the synthetic minority over-sampling technique (SMOTE) to generate new samples for minority classes. (4) Employ the attention-Fourier gate recurrent unit (AFGRU) classifier to capture long-term dependencies, with an adaptive weight assignment mechanism to enhance model generalization. **Results**: STANet achieves superior depression diagnostic performance, with 82.38% accuracy and a 90.72% AUC. The Spatio-Temporal Feature Aggregation module enhances classification by capturing deeper features at multiple scales. The AFGRU classifier, with adaptive weights and a stacked Gated Recurrent Unit (GRU), attains higher accuracy and AUC. SMOTE outperforms other oversampling methods. Additionally, spatio-temporal aggregated features achieve better performance compared to using only temporal or spatial features. **Comparison with existing methods**: STANet significantly outperforms traditional classifiers, deep learning classifiers, and functional connectivity-based classifiers. **Conclusions**: The successful performance of STANet contributes to enhancing the diagnosis and treatment assessment of depression in clinical settings on imbalanced and small fMRI.

## 1. Introduction

### 1.1. fMRI-Informed Depression Diagnosis

Depression is a global mental disorder that affects approximately 5% of the adult population, with a higher prevalence among women than men and senior adults than younger adults [1]. It is characterized by persistent low mood or reduced interest in activities, impacting emotions, cognition, and health, and serving as a risk factor for suicide [2]. The etiology of depression is multifactorial, encompassing genetic, environmental, psychological, and social factors. Depression can be categorized etiologically as endogenous or exogenous, reflecting different pathological mechanisms. Despite significant progress in the diagnosis and treatment of depression within psychiatry, its etiology and pathophysiology remain controversial and debated. Experienced psychiatrists can identify a wide range of depressive symptoms, including persistent sadness, loss of interest, changes in appetite and sleep, fatigue, difficulty concentrating, and thoughts of death or suicide. However, diagnostic criteria for depression vary across cultures and clinical practices, adding to the complexity of diagnosis [3,4]. Misdiagnoses can result in the adoption of improper treatment methods and the prescription of incorrect medications, worsening the depressive condition and posing a threat to the health of patients. Thus, developing a more reliable and precise diagnostic approach is essential and critical.

Neuroimaging studies in the past two decades have reported that patients with depression show atypical default mode network activity as measured by resting-state functional magnetic resonance imaging (rs-fMRI), and rs-fMRI measures are indicative of treatment effectiveness. Integrating behavioral measures with brain measures using a machine learning approach may provide better diagnosis and prognosis of depression, enhancing our understanding of the neurobiological mechanisms of depression and aiding in the accurate identification of depression and its subtypes [5,6]. Noman et al. [7] introduced graph autoencoder (GAE) and graph convolutional networks (GCN) in fMRI, learnt the embedding representation of the brain network using GAE, and identified depression by the learnt embedding. Lee et al. [3] utilized functional connectivity (FC) through a multispectral GCN and proposed a multispectral fusion framework for more reliable identification of major depressive disorder (MDD). Zhang et al. [8] proposed a deep residual contraction denoising network with channel-sharing soft thresholds for automatic depression identification. Additionally, Chen et al. [9] predicted depression using the amplitude of low-frequency (ALFF) and degree centrality (DC) of relevant brain regions, pinpointing abnormalities and providing insights into the underlying neural mechanisms.

### 1.2. fMRI-Informed Feature Integration

For a given time series, statistical domain features include histogram, interquartile range, mean absolute deviation, median absolute deviation, root mean square, standard deviation, and variance. Temporal domain features often encompass autocorrelation, centroid, mean absolute differences, distance, and entropy. Spectral domain features, derived from Fast Fourier Transform (FFT) or wavelet transformation (WT), include the FFT mean coefficient, wavelet absolute mean, wavelet standard deviation, wavelet variance, spectral distance, spectral entropy, wavelet entropy, and wavelet energy. Detailed expressions of these features are provided by Barandas et al. [10]. Considering the spatio-temporal properties of fMRI signals, integrating statistical, temporal, spectral, and spatial domain features simultaneously can significantly enhance depression diagnosis. Independent component analysis (ICA) is a commonly used method in fMRI data analysis, simultaneously extracting temporal and spatial features related to brain activity [11]. Moreover, methods such as amplitude of low frequency fluctuations (ALFF) [12], fractional ALFF (fALFF) [13], and spectrum contrast mapping (SCM) [14] are designed to map the spatial activity patterns in fMRI data through the spectral domain analysis. In recent years, the deep-learning-based fMRI feature integration has made great progress. For instance, Yan et al. [15] proposed a multi-scale recurrent neural network (RNN) model, which enabled classification schizophrenia and healthy controls by using time courses of fMRI-independent components directly [16]. Mao et al. [17] proposed an automatic diagnostic method using rs-fMRI data with spatio-temporal deep learning models based on granular computing. Liu et al. [18] proposed a spatial-temporal co-attention learning (STCAL) model for diagnosing ASD and ADHD which modeled the intermodal interactions of spatial and temporal signal patterns. Lee et al. [3] presented a multi-atlas fusion method that incorporates early and late fusion in a unified framework addressing the limitations that restricted their ability to capture the complex, multi-scale nature of the brain's functional networks. Lim et al. [19] showed a unified deep attentive spatio-spectral-temporal feature fusion framework to overcome the limitations of considering only a limited number of modes, which made it difficult to explore class-distinct spectral information of noise-related components.

### 1.3. Data Imbalance in fMRI-Based Classification Task

Data imbalance in machine learning arises when classes within a dataset are unevenly distributed, leading to biased model performance favoring the majority class and resulting in inaccurate predictions and misleading evaluation metrics for minority classes [20]. This challenge is particularly prevalent in neuroimaging studies, where data acquisition issues, such as subject absence during fMRI sessions, contribute to small and unbalanced sample sizes [16]. Various data augmentation techniques have been developed to address data imbalances and enhance classification models by expanding and balancing the dataset. Random oversampling, a straightforward approach involving the replication of minority class samples, has demonstrated effectiveness in several disease diagnostic applications [21]. However, its performance can be limited when applied indiscriminately across all samples. To mitigate these limitations, the synthetic minority over-sampling technique (SMOTE) was introduced, which synthesizes new minority class samples through interpolation [16,22–24]. For instance, Borderline-SMOTE is a modification of the classical SMOTE and is mainly used when the importance of the boundary samples is high and confusing [25]. SMOTE Tomek is a hybrid sampling technique that combines SMOTE with the Tomek Link removal method, which is suitable for datasets with significant noise and ambiguous boundaries [26]. SVMSMOTE integrates a support vector machine (SVM) with SMOTE to handle complex boundary structures and high-dimensional data [27]. Additionally, adaptive synthetic sampling (ADASYN) focuses on synthesizing minority class samples near decision boundaries, thereby enhancing model robustness [28]. These methods, particularly ADASYN and SMOTE, are widely applied in neuroimaging to improve minority class performance and overall model efficacy [16,23,24,29,30].

### 1.4. The Proposed Method

Based on the aforementioned considerations regarding intelligent depression diagnosis, we propose a novel Spatio-Temporal Aggregation Network (STANet) aimed at significantly improving the accuracy of depression diagnosis by addressing two key limitations in current diagnostic models: (1) The challenge posed by small and unbalanced fMRI samples; and (2) Inadequate integration of spatio-temporal features hindering effective fusion for depression diagnosis.

The remainder of this paper is organized as follows: Section 2 presents the dataset utilized, the preprocessing pipeline applied, and a detailed description of our proposed STANet. Section 3 includes a comparative analysis of performance against existing methods and ablation studies. Finally, Section 4 discusses the implications of our findings, including the advantages and limitations of STANet, in Discussion and Conclusion.

## 2. Materials and Methods

### 2.1. Dataset

The dataset was sourced from OpenNeuro (https://openneuro.org/) under accession number DS002748 [31]. It comprises 51 adult participants (13 Males and 38 Females) diagnosed with depression and 21 healthy controls (6 Males and 15 Females). Detailed demographic characteristics of the participants can be found in Bezmaternykh et al. [31]. Each session included 100 dynamic scans with 25 slices per brain volume. The resting-state fMRI scanning was conducted at the International Tomography Center, Novosibirsk, using a 3 T Ingenia scanner (Philips, Amsterdam, The Netherlands). Functional T2∗-weighted echo planar imaging scans were acquired using a fat suppression mode with voxel dimensions of 2 × 2 × 5 mm, a repetition time (TR) of 2500 ms, and an echo time (TE) of 35 ms. Participants were instructed to lie still with their eyes closed for 6 min. They gave their informed consent in accordance with the Helsinki Declaration and the ethics board of the Research Institute of Molecular Biology and Biophysics in Novosibirsk.

## 2.2. Pipeline of Data Processing

### 2.2.1. Pre-Processing

As illustrated in Figure 1, the data processing pipeline in this study comprises three sequential modules: Pre-processing, spatio-temporal feature aggregation, and classification. The pre-processing module is designed to preprocess the fMRI data following the standard pipeline using SPM12 software [32]. The initial five volumes of each scan were discarded to ensure data stability and temporal differences between slices within a volume were adjusted using the middlemost slice as the reference time point. No participant's scan had head movements exceeding 3 mm or head rotations exceeding 3°. All brain data were normalized to the Montreal Neurological Institute (MNI) space and smoothed with a Gaussian kernel of 8 mm.

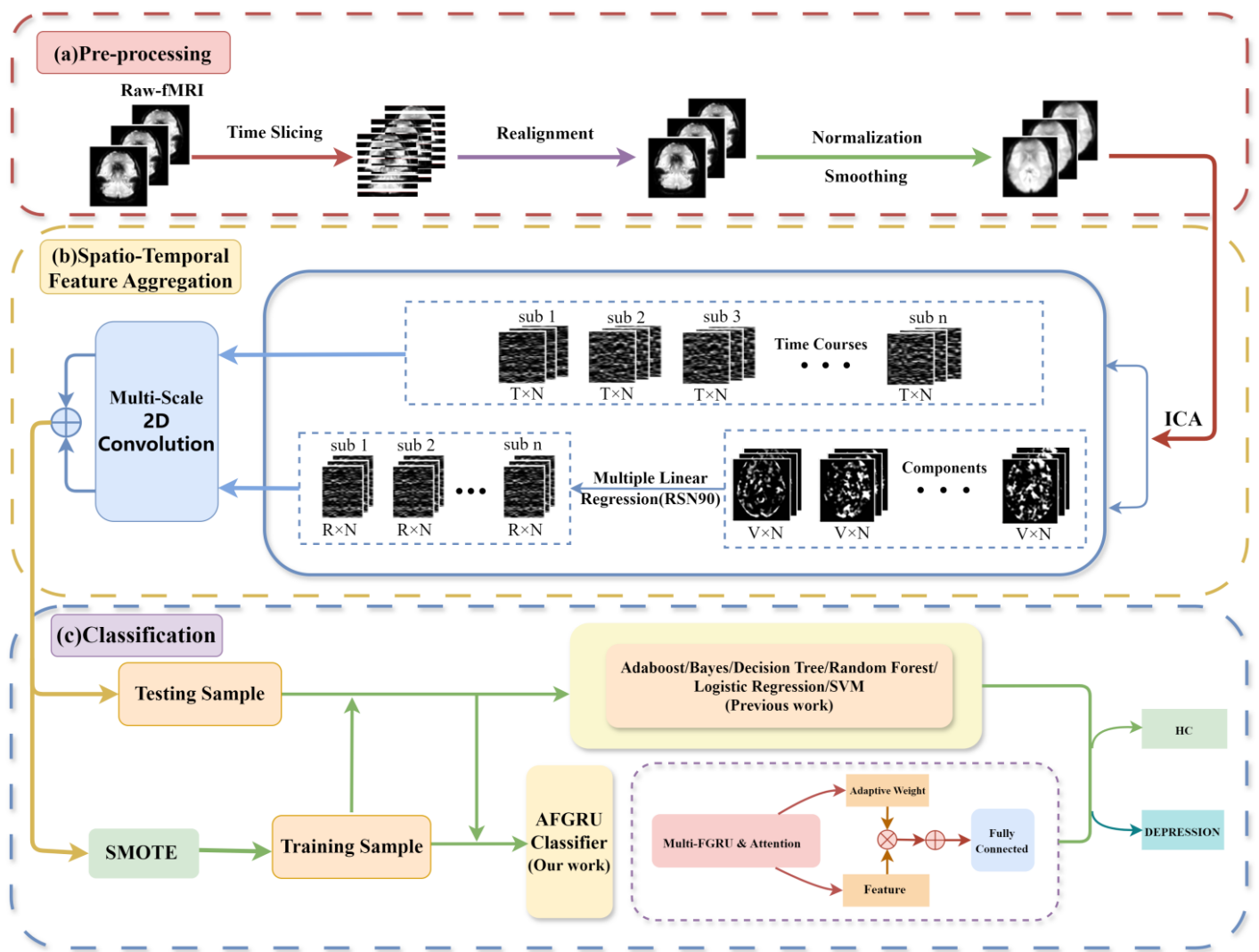

**Figure 1.** The flow diagram distinguishing depression patients from healthy controls. n: number of subjects; T: number of timepoints; N: number of source signals; V: voxel number of each spatial component; R: number of RSNs. (**a**) Data preprocessing. (**b**) Spatio-temporal feature aggregation:

Integration of time series features and spatial features. (**c**) Classification: Randomly divide the data into testing and training sets and use SMOTE for training sets, comparing the AFGRU Classifier with traditional machine learning methods (such as Adaboost, Bayes, Decision Tree, etc.) or previous work.

### 2.2.2. Model Architecture

In the Spatio-Temporal Feature Aggregation (STFA) module, we initially performed ICA on the preprocessed fMRI data to extract time courses and corresponding spatial maps. This was followed by multi-scale 2D convolution to form the fusion feature of spatio-temporal representation for each subject. Specifically, the GIFT tool [33] was employed to conduct Group ICA [34]. To obtain more stable independent components (ICs), we utilized ICASSO [35] for the analysis, ultimately selecting 17 ICs based on the optimal estimation of order number [36]. Furthermore, multiple linear regression was applied to the time courses and spatial map features obtained by the ICA to determine the spatial similarity with the resting-state network (RSN) atlas [37].

With regard to the classification module, the fusion features of spatio-temporal representation generated by the STFA module for each subject are fed into various classifiers to perform the depression classification task. Specifically, in the training stage, SMOTE is applied to address the imbalance in fMRI samples.

### 2.3. STANet

The detailed structure of the proposed STANet is illustrated in Figure 2. It primarily comprises three components: STFA, SMOTE, and the AFGRU classifier. The STFA module is responsible for generating the fusion features of spatio-temporal representation. SMOTE is employed to address the issue of data imbalance. The AFGRU classifier is designed to enhance classification performance on the small-sized depression dataset.

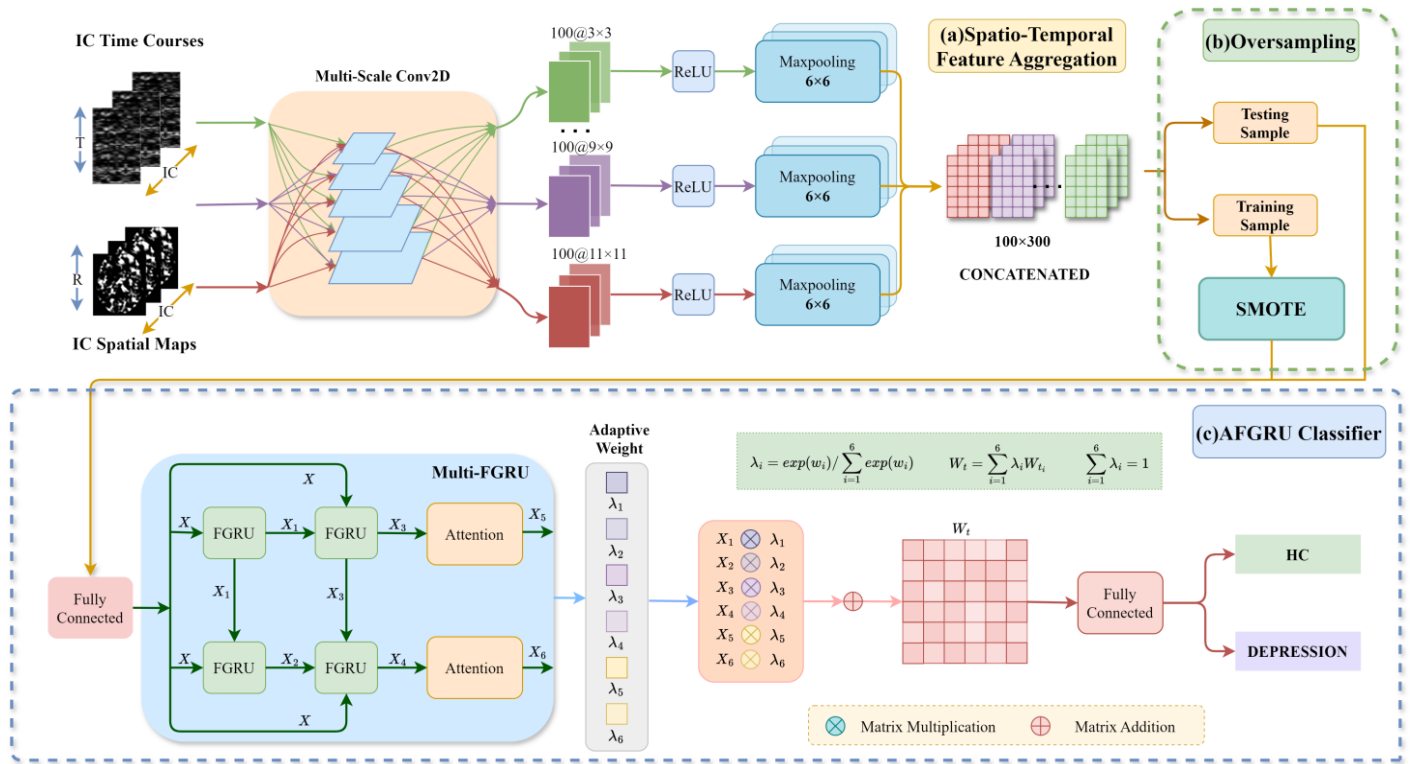

**Figure 2.** The detailed structure of STANet. (**a**) Spatio-Temporal Feature Aggregation: The pre-processed data undergoes ICA to extract the independent component (IC) time courses and IC spatial maps, and then the resulting spatial map features are subjected to multiple linear regression, pooled, and concatenated with the time series before being fed into the next module. (**b**) Oversampling: The data are randomly divided into training and testing sets. The SMOTE is applied to the training set to balance the dataset. (c) AFGRU Classifier: The extracted features are input into the Multi-FGRU. Features obtained at each stage are assigned adaptive weights, and classification performance is evaluated using a 10-fold cross-validation strategy.

### 2.3.1. STFA Module

STFA intelligently integrates multi-scale spatio-temporal information. Specifically, STANet obtains time series and spatial features through ICA, then integrates the spatial features with RSN for multiple linear regression, followed by a multi-scale spatio-temporal integration. This selective mechanism allows for a more targeted and efficient use of spatio-temporal features, potentially leading to better model performance.

#### Independent Component Analysis

Independent Component Analysis (ICA) is a widely-used technique to extract independent features from high dimensional fMRI data. The core principle of ICA is to decompose the observed mixed data into statistically and spatially independent components and their associated time courses [11,38–42]. Let $X$ denote a single subject's fMRI data with T time points and V voxels within brain. Here, $S$ is an N × V matrix containing N source signals, which are assumed unobservable, mutually statistically independent, and non-Gaussian. Each row represents an independent component (IC). Furthermore, $A$ is a T × N unknown mixing matrix that contains the associated time courses of N source signals. Consequently, the ICA model can be represented as:

$$X = AS. \tag{1}$$

The objective of solving ICA is to estimate an N × T matrix $W$, such that Y is a good approximation of the source signals S by the following formula:

$$Y = WX. \tag{2}$$

#### Multiple Linear Regression

To capture the implicit relation between ICs Y generated by ICA in Formula (2), the ICs were then mapped to a RSN template [37] to perform multiple linear regression. This process results in a matrix representing spatial similarity features. The multiple linear regression formula can be expressed as:

$$Q = Y\beta, \tag{3}$$

where Y represents the estimated source signals, $Q$ denotes the spatial similarity matrix between the estimated source signals and the RSN template with dimensions N × R, and $\beta$ is the regression coefficient matrix with dimensions V × R.

#### Multi-Scale Convolution Layer

CNNs are highly effective at processing image data, particularly for extracting spatial features. Given the spatial nature of fMRI data, we utilized CNNs to accurately identify and extract key regions of brain activity.

To integrate local information, we employed multi-scale 2D convolutional layers using five different scales of 2D convolution kernels. This approach facilitates comprehensive feature extraction and efficient utilization of the available space for information extraction. We utilized convolutional kernels of varying sizes (3 × 3, 5 × 5, 7 × 7, 9 × 9, 11 × 11) to ensure diversity and comprehensiveness. To address the potential presence of negative values during convolution, we incorporated a ReLU layer to maintain stability and effectiveness in parameter learning. Subsequently, a 6 × 6 max-pooling layer was applied for downsampling along the time dimension, resulting in feature representations of uniform size for both time courses and spatial components. To further enhance the feature representation, we concatenated the features obtained from the convolution kernels at each scale for both time courses and spatial components. This final feature representation, achieved through the concatenation layer, provides richer and more precise inputs for subsequent oversampling methods.

### 2.3.2. SMOTE

Due to the complexity and specificity of the fMRI, obtaining a sufficiently large number of subjects is often challenging, resulting in small and unbalanced datasets. Directly feeding such data into the classifier can cause it to overlearn from the majority class, skewing the test results. To mitigate this issue, we employ SMOTE to process the training dataset, synthesizing minority class data to achieve a balanced dataset. The balanced training set is then used to train the classifiers. We opted to employ SMOTE primarily based on its ability to generate new synthetic samples by interpolating between minority class instances without introducing noise, effectively addressing the issue of data imbalance. Compared to other methods, SMOTE preserves the distribution characteristics of the dataset, thus reducing the risk of overfitting. Furthermore, the widespread use and robust performance of SMOTE in numerous studies [16,23,24] underscore its effectiveness and reliability in dealing with imbalanced datasets. Consequently, in this study, we utilize SMOTE and achieved optimal results. Figure 3 illustrates the data distribution before and after SMOTE processing, which successfully generates an approximately balanced training dataset.

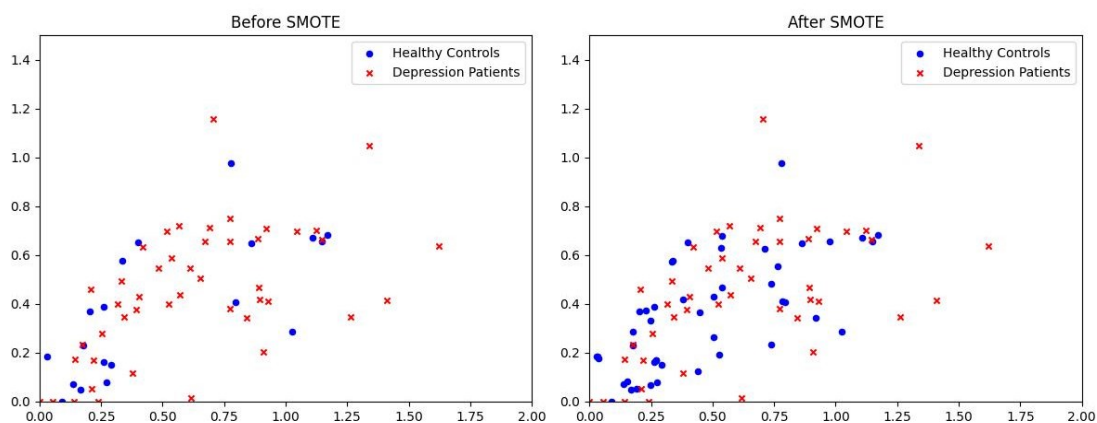

**Figure 3.** Distribution of positive and negative samples in the training dataset: (**left**) without SMOTE, (**right**) with SMOTE.

### 2.3.3. AFGRU Classifier

In the realm of time-series prediction and sequence modeling, selecting between LSTM and GRU architectures is crucial [43]. Both are recurrent neural network variants designed to address the vanishing gradient problem and capture long-term dependencies, yet they possess distinct characteristics. The AFGRU classifier, which integrates stacked GRUs with Fourier transform capabilities, exemplifies this comparison:

LSTM networks are renowned for their robustness in capturing long-term dependencies due to their complex gate structure. As a more recent innovation, GRU is simpler and more computationally efficient than LSTM. This streamlined structure allows GRUs to train faster and often requires fewer parameters, which is advantageous for datasets with limited size or computational resources. In the case of the AFGRU classifier, the choice of GRU is driven by the need for a more computationally efficient model that can still capture complex temporal patterns. The incorporation of Fourier transform within the GRU framework enables the AFGRU Classifier to analyze both the time and frequency domains. Additionally, the adaptive weighting mechanism applied to the outputs of the stacked GRUs is a key feature of the AFGRU Classifier. This mechanism allows the network to dynamically allocate weights to the features extracted at each time step, fine-tuning the importance of different temporal intervals in the prediction process. This adaptability is particularly beneficial in scenarios where the significance of time intervals can vary, as is often the case in real-world time-series data. Consequently, the AFGRU Classifier was deemed the most suitable for our study.

Multi-FGRU

Considering the temporal features of fMRI in the latent space, we developed the AF-GRU Classifier. RNNs capture temporal correlations and incorporate historical information, which is essential for fMRI data analysis [43]. We employed GRU [44], a robust mechanism within RNNs. By stacking multiple GRU layers, we effectively address the issues of gradient explosion and gradient vanishing, thereby enhancing the model's representation and learning capabilities to capture higher-level dynamic information. The integration of GRU into the data processing flow allows for controlled information accumulation [45], including selective addition of new information and selective forgetting of previously accumulated information, with the hidden layer size set to 200. To further augment the model's ability to process complex neural signals, we incorporated the fast Fourier transform (FFT) into the GRU model [46]. The FFT converts time-domain signals into frequency-domain signals, enabling the extraction and analysis of characteristic information from different frequency components. The combination of frequency-domain features and time-domain features enables the model to more comprehensively understand and model the complex activity patterns of the brain. Following the FGRU layer, we introduce an attention mechanism layer to help the model focus on the most relevant parts of the input sequence. During data processing, the information processed by the FGRU layer may gradually degrade. However, the introduction of the attention mechanism enables us to better capture the important features of different parts of the sequence, thereby reducing information loss. This enhancement allows the model to more effectively handle long sequential data and capture long-distance dependencies within the sequences.

Firstly, the proposed FGRU involves applying the FFT operation to the input data as follows:

$$x_{fft} = \text{Real}(FFT(x_t)), \tag{4}$$

where $x_t$ denotes current moment input information, $FFT(\cdot)$ represents FFT operation, and $Real(\cdot)$ means the extraction of the real part. Then, the GRU involves three main gating processes: update gate, reset gate, and update the hidden state. The GRU unit update gate and reset gate are expressed as:

$$z_t = \sigma(\boldsymbol{W}_z \cdot [h_{t-1}, x_{fft}]), \tag{5}$$

$$r_t = \sigma(\boldsymbol{W}_r \cdot [h_{t-1}, x_{fft}]), \tag{6}$$

where $\boldsymbol{W}_z$ denotes the weight matrix of the update gate, $\boldsymbol{W}_r$ represents the weight matrix of the reset gate, $h_{t-1}$ is hidden state at the previous moment, $\sigma$ denotes the sigmoid function, $r_t$ denotes the reset gate, and $z_t$ is the update gate. Meanwhile, the candidate hidden state and the final hidden state are computed as:

$$\widetilde{h_t} = tanh(\boldsymbol{W} \cdot [r_t * h_{t-1}, x_{fft}]), \tag{7}$$

$$h_t = (1 - z_t) * h_{t-1} + z_t * \widetilde{h_t}, \tag{8}$$

where $\boldsymbol{W}$ denotes the weight matrix of the hidden layer, $h_t$ is the hidden state passed to the next moment, $\widetilde{h_t}$ is the candidate hidden state, and $tanh$ is the hyperbolic tangent function.

These formulas describe the computational process for a single FGRU. In the AFGRU classifier, each FGRU transforms $h_t$ from the previous step to generate deeper feature representation. In calculating $h_t$, $\widetilde{h_t}$ is used to balance retained and updated information.

Combined with $z_t$, the computation of $h_t$ retains old information while integrating new information, allowing the model to make appropriate updates and adjustments as it processes sequence data. Through these gating mechanisms, the FGRU selectively passes information to the next time step, learning long-term dependencies more efficiently. The AFGRU classifier leverages the strengths of both frequency domain transformations and RNNs to identify and preserve intricate temporal patterns, enabling the model to learn complex underlying time series patterns and features.

Adaptive Weighting

Adaptive weighting is extensively used in signal processing, machine learning, image processing, and other fields. It helps models better adapt to data characteristics during training, thereby improving accuracy and generalization. In this study, initial weights are set randomly, and 500 rounds of weight updates are performed. Adaptive weights are assigned to X1, X2, X3, X4, X5, and X6 in the AFGRU Classifier (Figure 2) to enhance the model's generalization ability by treating each data step as part of feature processing. The process of adaptive weighting is detailed in Algorithm 1.

---

**Algorithm 1**: Adaptive Weighting

**Input**: Sample data $(x_i, y_i)$, sample data weights $w_i$, training iteration number $L_i (0 < i \leq 6)$
**Output**: Optimal model
**Initialization**: Set $w_i$ to Gaussian distribution random number and $\sum_{i=1}^{6} w_i = 1$
**Start**:

    For i from 0 to $L_i$:
        #Train the model using the current weights
        model = Train $((x_i, y_i), w_i)$
         #Calculate the loss function
         Loss = MSE $(model, (x_i, y_i))$
         #Update sample weights to minimize the loss function
        For j = 1 to 6:
            Prediction value = model. predict $(x_i)$
            Truth value = $y_i$
            $w_i = w_i * \exp(-\text{lr} * (\text{Prediction value} - \text{Truth value}))$
        End for
        #Normalize sample weights
        For k = 1 to 6:
            $w_i = w_i / \sum_{i=1}^{6} w_i$
        End for
    End for
Return

---

*2.4. Performance Metrics*

We employ four metrics—accuracy (ACC), F1-score, recall (Recall), and area under the curve (AUC)—to evaluate the performance of STANet in classifying depression and normal controls. The corresponding formulations are defined as follows:

$$ACC = \frac{TP+TN}{TP+TN+FP+FN}, \tag{9}$$

$$SEN = \frac{TP}{TP+FN}, \ PPV = \frac{TP}{TP+FP}, \tag{10}$$

$$F1-score = 2 * \frac{SEN \times PPV}{SEN+PPV}, \tag{11}$$

$$Recall = \frac{TP}{TP+FN},\qquad(12)$$

where TP, TN, FP, FN, and PPV stand for true positive, true negative, false positive, false negative, and positive predictive value, respectively.

In clinical research and decision-making, evaluating key performance metrics for diagnostic tests or predictive models is crucial for optimizing patient care. The recall rate is essential for ensuring that patients who need treatment are identified, thereby reducing the risk of severe outcomes. A high recall rate ensures timely intervention for most patients requiring treatment. The F1-score offers a balanced assessment by combining recall and precision, aiding clinicians in achieving an appropriate balance in diagnostic and treatment decisions. The AUC reflects the model's ability to distinguish between patients and non-patients, with a high AUC indicating greater reliability in clinical decision support. This facilitates the development of more precise treatment plans. Therefore, a comprehensive consideration of these metrics is vital for enhancing the accuracy of clinical diagnoses and the efficacy of treatments.

## 3. Results

### 3.1. Experimental Setting

We employ a ten-fold cross-validation strategy at the subject level to evaluate the performance of STANet. Specifically, all the subjects are evenly divided into ten sets. One set is used as the test set, while the remaining nine sets are used for training. This process is repeated ten times, allowing each set to be used for testing in turn. In Figure 1, the preprocessing excludes the first five time points, leaving 95. The ICs are processed by GIFT to determine the optimal number of 17 automatically. For the spatial components obtained after ICA processing, multiple linear regression is performed with the RSN template [37], resulting in spatial features of 90 × 17.

The training and classification of the classifiers in this study were conducted on an Nvidia GeForce GTX3060 GPU with 12GB RAM, using classification models written in Python 3.8 on a Windows 10 environment. During STANet training, the MSE loss function was used with a learning rate empirically set to 0.01.

### 3.2. Performance Assessment of STFA Module in STANet

#### 3.2.1. Performance Comparison Without STFA Module

We compared STANet with five traditional popular classifiers (decision tree (DT), SVM, random forest (RF), Adaboost, logistic regression (LR)) and six RNN-based deep learning models, all tested using ten-fold cross-validation. Among conventional classifiers, SVM performs the best, with a classification accuracy of 66.61%. In contrast, STANet achieved a classification accuracy of 82.38% and an AUC of 90.72%, significantly outperforming Adaboost, RF, LR, DT, and SVM. Comparisons with other GRU-based RNN structures further verified the advantages of the proposed model, showing improvements in classification accuracy.

Table 1 demonstrates the classification performance of the six methods using time courses and spatial components of processed independent components as training data, with SMOTE applied beforehand. Notably, STANet achieved a classification accuracy of 82.38%, while traditional classifiers like SVM and RF had accuracies of around 65%, significantly lower than STANet. The performance of individual GRU or LSTM models was suboptimal. Combining a GRU layer with a 2D convolutional layer enhanced classification performance. Therefore, our proposed STANet leverages the strengths of CNN and RNN to learn both temporal and spatial features, and incorporates adaptive weights to improve generalization, achieving the best performance.

**Table 1.** Classification performance comparison without the STFA module among competing methods.

| Methods | Accuracy | F1-Score | Recall | AUC |
|---------|----------|----------|--------|-----|
| Adaboost | 51.25% | 63.47% | 62.33% | 43.67% |
| DT | 52.86% | 64.74% | 64.67% | 44.83% |
| GRU | 52.68% | 60.19% | 55.33% | 45.50% |
| LSTM | 47.32% | 55.41% | 51.33% | 48.17% |
| LG | 65.54% | 78.69% | 92.33% | 51.33% |
| RF | 63.75% | 77.20% | 90.00% | 51.17% |
| SVM | 66.61% | 79.54% | 94.00% | 50.33% |
| STANet | 82.38% | 88.18% | 82.38% | 90.72% |

### 3.2.2. Performance Comparison with STFA Module

Comparison of Tables 1 and 2 reveals that data processed through STFA, followed by classification using traditional classifiers, achieve higher ACC and AUC. This highlights the importance of multi-scale convolution in data processing. The improved performance metrics underscore STFA's ability to effectively capture diverse features and patterns, leading to more accurate and reliable classification outcomes. Transformers was added for comparison, but it did not work well and made the model too bloated.

**Table 2.** Classification performance comparison with the STFA module among competing methods.

| Method | Accuracy | F1-Score | Recall | AUC |
|--------|----------|----------|--------|-----|
| STFA-Adaboost | 77.86% | 83.31% | 82.67% | 74.67% |
| STFA-DT | 76.61% | 82.39% | 82.67% | 72.17% |
| STFA-GRU | 63.93% | 73.03% | 74.67% | 52.83% |
| STFA-LSTM | 43.04% | 41.82% | 44.00% | 49.67% |
| STFA-LG | 75.18% | 83.30% | 88.33% | 75.83% |
| STFA-SVM | 67.14% | 72.22% | 82.17% | 28.42% |
| STFA-RF | 68.21% | 77.45% | 80.67% | 79.67% |
| STFA-Transformer | 72.38% | 82.21% | 75.86% | 83.72% |
| STANet | 82.38% | 88.18% | 82.38% | 90.72% |

### 3.3. Performance Assessment of AFGRU Classifier in STANet

To verify the advantages of the AFGRU Classifier in STANet, we compared it with other GRU-based RNN models. As shown in Table 3, the single LSTM achieved an accuracy of only 43%, while the single GRU reached 63%, indicating that GRU outperforms LSTM in both ACC and AUC, whereas the stacked GRU model showed a clear advantage. After processing the stacked GRU modules, we introduced an attention mechanism and assigned adaptive weights to enhance the model generalization. Simply stacking GRU layers improved accuracy to 66.67%. Introducing Fourier transforms and processing the frequency domain of the data, the STFA-AtFGRU model increased accuracy to 73.49%, and the STFA-AdFGRU model achieved 76.34%. Comprehensive processing with STFA-AFGRU further increased ACC to 82.38% and AUC to 90.72%. Table 3 demonstrates that incorporating Fourier transforms into GRU significantly improves ACC and AUC by leveraging frequency domain information, enriching the model's ability to capture complex temporal dependencies. The AFGRU classifier's superior performance underscores the benefit of integrating Fourier transforms for advanced sequence modeling.

**Table 3.** Ablation performance comparison of STANet with regard to AFGRU classifier.

| Methods | Accuracy | F1-Score | Recall | AUC |
|---------|----------|----------|--------|-----|
| STFA-sLSTM | 43.04% | 41.82% | 44.00% | 49.67% |
| STFA-sGRU | 63.93% | 73.03% | 74.67% | 52.83% |

| | | | |
|---|---|---|---|
| STFA-dGRU | 66.67% | 71.54% | 69.76% | 77.72% |
| STFA-AtFGRU | 73.49% | 81.26% | 82.33% | 86.33% |
| STFA-AdFGRU | 76.34% | 84.03% | 79.17% | 87.11% |
| STFA(s)-AFGRU | 77.78% | 85.19% | 80.40% | 74.78% |
| STFA-AGRU | 79.52% | 86.24% | 81.81% | 89.72% |
| STANet(t) | 66.67% | 77.76% | 69.81% | 46.50% |
| STANet(s) | 73.81% | 82.84% | 77.67% | 81.44% |
| STANet | 82.38% | 88.18% | 82.38% | 90.72% |

Notes: STFA: Spatio-temporal feature aggregation. STFA(s): Spatio-temporal feature aggregation (single-CNN), is only convolutional kernel is 7 × 7 convolution. sLSTM: single-LSTM, only one layer of LSTM is used for classification after the convolutional layer. sGRU: single-GRU, only one layer of GRU is used for classification. dGRU: double-GRU, double layers of GRU are used for classification. AtFGRU: AFGRU classifier without adaptive mechanism layer. AdFGRU: AFGRU classifier without attention mechanism layer. AGRU: AFGRU classifier without Fourier transform. STANet(t) is STANet with only temporal information as input. STANet(s) is STANet with only spatial information as input.

In terms of convolutional layers, multiple scales are superior to single convolution, as they capture richer data. Our proposed STANet assigns adaptive weights to data from the GRU module, achieving optimal performance. This co-training approach enhances convolutional visual representations and temporal dynamics, leading to better results.

By comparing Tables 2 and 3, it can be seen that deep learning can reach a higher accuracy compared to traditional classification models, and it also confirms that the combination of CNN and RNN can obtain a higher performance for classification of fMRI data.

STANet(t) and STANet(s) illustrate the importance of input type. Our model, which combines time series and spatial regression inputs, significantly outperforms models using either input alone. This synergy enhances the model's overall accuracy and robustness.

### 3.4. Oversampling Strategy Impact on STANet

To compare the effects of different data balancing methods on classification performance, we used six methods: Random oversampling, SMOTE, ADASYN, borderline-SMOTE, SMOTE Tomek and SVMSMOTE. The classification results, shown in Table 4, indicate that SMOTE significantly outperforms the other methods, achieving the highest AUC. This suggests that the data generated by SMOTE is more consistent with the original data distribution than the other methods.

**Table 4.** Performance comparison among different oversampling strategies adopted in STANet.

| Method | Accuracy | F1-Score | Recall | AUC |
|---|---|---|---|---|
| Random Oversampling | 76.67% | 84.53% | 78.38% | 81.06% |
| SMOTE | 82.38% | 88.18% | 82.38% | 90.72% |
| ADASYN | 75. 24% | 82. 04% | 85.14% | 86.39% |
| Borderline-SMOTE | 78.10% | 85.75% | 79.52% | 85.39% |
| SMOTE Tomek | 74.92% | 83.58% | 79.52% | 88.06% |
| SVMSMOTE | 72.38% | 81.56% | 75.10% | 80.00% |

### 3.5. Order Number Impact on STANet

To compare the effect of different numbers of ICs on classification performance, we manually set the number of ICs to 15, 21, 24, and 27, with 17 as the best estimated number for comparison. The classification results, shown in Table 5, indicate that the best performance and highest AUC are achieved when the number of ICs is set to the best estimated value.

**Table 5.** Classification performance of STANet under different order numbers in ICA decomposition.

| Number of ICs | Accuracy | F1-Score | Recall | AUC |
|---|---|---|---|---|
| 15 | 72.38% | 82.62% | 74.81% | 63.78% |
| 17 (estimated) | 82.38% | 88.18% | 82.38% | 90.72% |
| 21 | 68.10% | 80.34% | 69.76% | 63.33% |
| 24 | 63.81% | 76.18% | 69.00% | 60.00% |
| 27 | 69.52% | 81.15% | 73.71% | 66.61% |

Notes: Figures A1–A5 display the spatial maps obtained by ICA decomposition with varying order numbers. Order number 17 is automatically estimated by GIFT v3.0b software.

### 3.6. Comparison with Other Competing Methods

As shown in Table 6, the proposed STANet significantly outperforms other state-of-the-art models, suggesting that our model has the potential to aid in the diagnosis of depression. Both Convolution-GRU [16] and Auto-ASD-Network [23] were chosen to balance the dataset using SMOTE, and Co-Teaching Learning [47] has also been effective for fMRI-based diagnosis of depression. Models like MsRNN [15], Spectral-GNN [3], and wck-CNN [48] achieved only about 70% accuracy, indicating that STANet has superior performance on the imbalanced depression dataset. In the realm of spatio-temporal modeling, while models such as STCAL [18] and STGCN [49] have their own merits, STANet also has demonstrated certain merits. The AFGRU classifier as part of the STANet dynamically adjusts the contribution of each layer or model. This adaptive mechanism enables the model to focus more intently on salient information, thereby enhancing its ability to capture and balance both long-term and short-term dependencies.

**Table 6.** Classification performance comparison among different competing models.

| Method | Input | Accuracy | F1-Score | Recall |
|---|---|---|---|---|
| Convolution-GRU | Time Courses | 65.24% | 77.58% | 69.24% |
| Auto-ASD-Network | Time Courses | 75.24% | 83.67% | 79.57% |
| MsRNN | Time Courses | 73.81% | 82.72% | 76.48% |
| Co-Teaching Learning | FC Matrix | 70.95% | 79.40% | 79.19% |
| Spectral-GNN | FC Matrix | 69.59% | 70.07% | 68.99% |
| wck-CNN | FC Matrix | 63.04% | 59.84% | 58.69% |
| STCAL | Spatio-Temporal | 76.67% | 84.75% | 79.19% |
| STANet | Spatio-Temporal | 82.38% | 88.18% | 82.38% |

In 2022, Dai et al. [50] and Chen et al. [9] trained and validated their models using the same dataset, achieving an accuracy of 68.9% and an AUC of 89.4%. These results indicate that STANet significantly outperformed other studies in terms of performance.

## 4. Discussion

### 4.1. Performance Analysis

For a long time, the diagnosis of depression has primarily relied on a comprehensive assessment of clinical symptoms. Recently, numerous studies have attempted to identify stable fMRI-based biomarkers using machine learning techniques. In this study, to further diagnose depression, we employed the ICA method to extract independent components. The resulting time courses and spatial components were integrated using STFA. We then applied the SMOTE method to balance the training set by adjusting the number of minority samples. The AFGRU classifier was utilized to extract potential information from the temporal dimension of the data. Finally, adaptive weighting was employed to enhance the model's ability to handle new samples. This approach achieved an accuracy of 82.38% and an AUC of 90.72%, representing a 5% improvement in accuracy compared to traditional methods. These results indicate a significant enhancement in the predictive discrimination ability of deep learning in neuroimaging.

In this study, we employed traditional classifiers such as SVM, DT, RF, and LG. However, the results indicate that these traditional classifiers performed poorly. This may be attributed to the high feature dimensions and strong nonlinearity present in the data, which adversely affect classification performance. In contrast, classifiers such as SVM and LG are essentially linear classifiers with stringent data requirements. Compared to other deep learning methods, it further demonstrated the superior performance of STANet. Additionally, the FC matrix [51] was used as an input for classification in the neuroimaging field, as shown in Table A1. The results clearly indicate that the classification performance using the FC matrix is inferior to that obtained using ICs as input.

### 4.2. Diagnostic Analysis of Depression

Several studies have shown that the diagnosis of depression is related to the frontal lobe [52], parietal lobe, temporal hippocampus, and amygdala [51], among others. Frontal lobe trauma may lead to executive function deficits, decision-making difficulties, and difficulties in emotion regulation; the temporal and parietal lobes have been associated with memory problems, language deficits, and difficulties in spatial perception; the amygdala and the hippocampus [53] are closely linked functionally and work together to process and remember emotionally relevant information. Similar results were observed in the ICA results of fMRI in this study, which also proves that ICA is a good tool for studying brain patterns. From previous studies, we can learn indirectly through classifiers or statistical methods, whereas ICA provides us with the opportunity to directly study the independent components of brain activity in combination with classification methods. In terms of the proposed STANet method, it may be beneficial in identifying brain network features associated with different subtypes of depression, revealing neurobiological differences between these subtypes, and enhancing clinical diagnostic accuracy. By objectively measuring biomarkers of brain activities, this method likely provides quantitative indicators for diagnosis and treatment evaluation. This helps monitor disease progression and treatment effects, offering a scientific basis for treatment adjustment.

### 4.3. Limitation and Future Work

Regarding the proposed STANet, specific values are not assigned to the hidden states of each GRU, which likely enhances performance by incorporating a weighted mechanism within the GRU in the future. However, we recognize that the current tools may not be readily accessible to clinicians. To address this gap, we are actively implementing a dedicated toolbox tailored to the requirements of clinicians or psychologists, which will be accessible publicly in future. The homogeneity and heterogeneity of the diagnostic structures were not analyzed in depth in this study, potentially affecting the generalizability and objectivity of the results. Future studies should consider these factors to enhance the robustness of the assessment. Furthermore, structural brain imaging plays a pivotal role in depression research, uncovering critical anatomical alterations which are associated with disease symptoms and cognitive impairments [54–56]. In future work, we aim to integrate structural MRI with fMRI, providing a more comprehensive set of biomarkers for the diagnosis and treatment of depression. Meanwhile, we will address issues such as sample size, population diversity, and fMRI data biases by establishing a multi-center collaboration project. This initiative will highlight the impact of these factors on the generalizability and performance of our model.

## 5. Conclusions

In this study, we proposed STANet for diagnosing depressive disorder, integrating CNN and RNN to capture both temporal and spatial features of brain activity, which includes spatio-temporal feature aggregation, multi-scale deep convolution, data balancing with SMOTE, and the AFGRU classifier with an assignment of adaptive weights. Experimental results demonstrate that STANet achieves superior diagnostic performance and outperforms traditional classifiers, deep learning classifiers, and functional connectivity-based classifiers. This approach provides a robust framework for leveraging fMRI and artificial intelligence to improve the accuracy and reliability of depression diagnosis. Our diagnostic approach is expected to improve the recognition rate of depression based on brain fMRI scanning, thereby facilitating timely treatment and improving patient prognosis. Additionally, by optimizing the diagnostic process, we can reduce misdiagnoses and missed diagnoses, thereby improving the effective use of medical resources.

**Author Contributions:** Conceptualization, W.Z. (Wei Zhang), W.Z. (Weiming Zeng) and J.L.; methodology, W.Z. (Wei Zhang), W.Z. (Weiming Zeng), J.L. and N.W.; software, W.Z. (Wei Zhang) and H.C.; validation, W.Z. (Wei Zhang), H.Y., K.Z., R.T., W.T.S. and N.W.; formal analysis, W.Z. (Weiming Zeng) and N.W.; investigation, W.Z. (Wei Zhang) and N.W.; resources, H.Y.; data curation, W.Z. (Weiming Zeng) and H.C.; writing—original draft preparation, W.Z. (Wei Zhang); writing—review and editing, W.Z. (Wei Zhang), H.Y., K.Z., R.T., W.T.S. and N.W.; visualization, H.C.; supervision, W.Z. (Weiming Zeng) and N.W.; project administration, N.W.; funding acquisition, W.Z. (Weiming Zeng), K.Z. and N.W. All authors have read and agreed to the published version of the manuscript.

**Funding:** This research was funded by the National Natural Science Foundation of China [grant numbers: 31870979; 12304526], the Hong Kong Polytechnic University Faculty Reserve Fund (Project ID: P0053738), and The Hong Kong Polytechnic University Start-up Fund (Project ID: P0053210).

**Institutional Review Board Statement:** Not applicable

**Informed Consent Statement:** Not applicable.

**Data Availability Statement:** The data presented in this study are in a publicly accessible repository and openly available in OpenNeuro (https://openneuro.org/) under Accession Number DS002748 (Bezmaternykh et al., 2021)

**Conflicts of Interest:** The authors have no conflicts of interests to disclose.

### Appendix A. Traditional Classifiers Based on the FC Matrix

Considering that the functional connectivity (FC) matrix is often used as an input for a classification task, we obtained the FC matrix of the subjects based on the AAL116 template [57] for performance comparison. Given that the FC matrix is not time series data, we selected classical classifiers, all trained using ten-fold cross-validation. As shown in Table A1, Bayes achieved the highest classification accuracy of 63.75%. The AUC only reached 48.33%, slightly lower than 55.00% AUC of DT, but with more balanced performance. This suggests that the traditional classifiers represented by SVM and Bayes are more advantageous in the classification of the FC matrix. However, the accuracy and AUC of traditional classifiers did not exceed 60%. In contrast, the deep learning model demonstrated superior classification performance in this study, while most traditional classifiers performed poorly. This indicates that the deep learning model is more suitable for the classification task in this study compared to traditional classification models.

**Table A1.** Classification performance comparison among traditional classifiers based on the FC matrix.

| Methods | Accuracy | F1-Score | Recall | AUC |
|---------|----------|----------|--------|-----|
| Adaboost | 52.50% | 61.43% | 59.00% | 48.67% |
| Bayes | 63.75% | 76.12% | 84.33% | 48.33% |
| DT | 59.46% | 66.81% | 66.67% | 55.00% |
| RF | 62.14% | 75.11% | 84.00% | 40.33% |
| LG | 51.25% | 65.12% | 68.33% | 32.00% |
| SVM | 60.89% | 73.92% | 65.65% | 47.17% |

Notes: The Pearson correlation coefficients between each pair of brain regions were calculated using the AAL116 template, resulting in a functional connectivity matrix as input for this dataset.

## Appendix B. Results of Group ICA

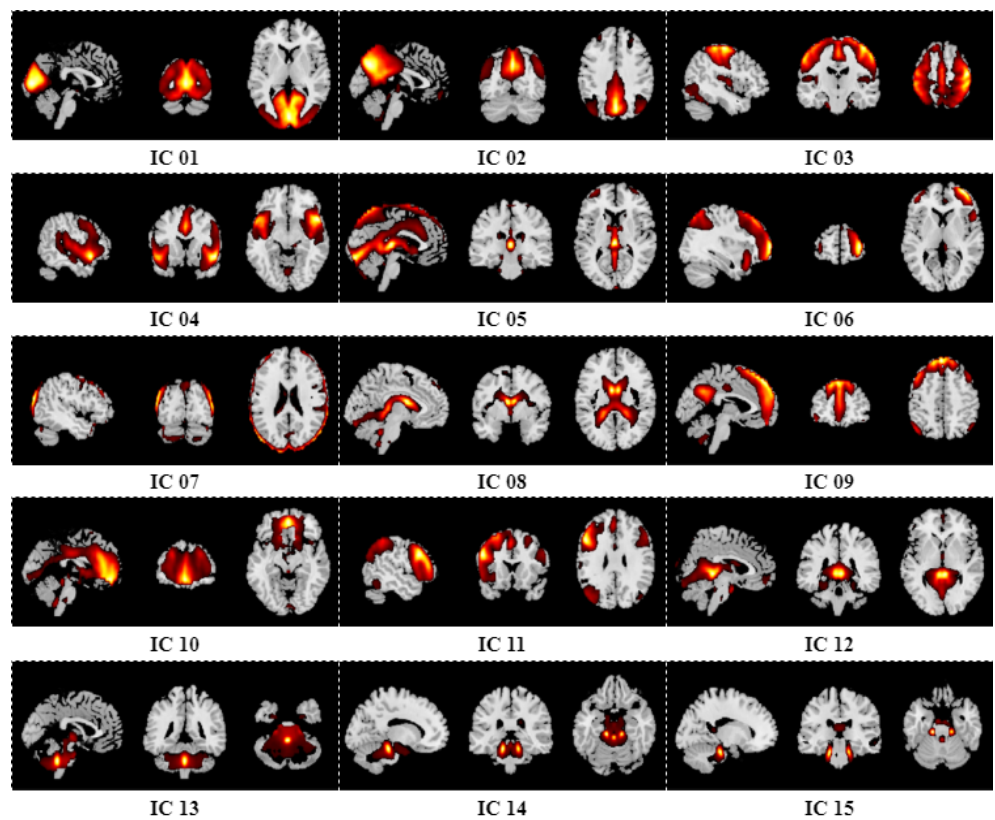

**Figure A1.** Results of spatial maps after group ICA with order number equal to 15.

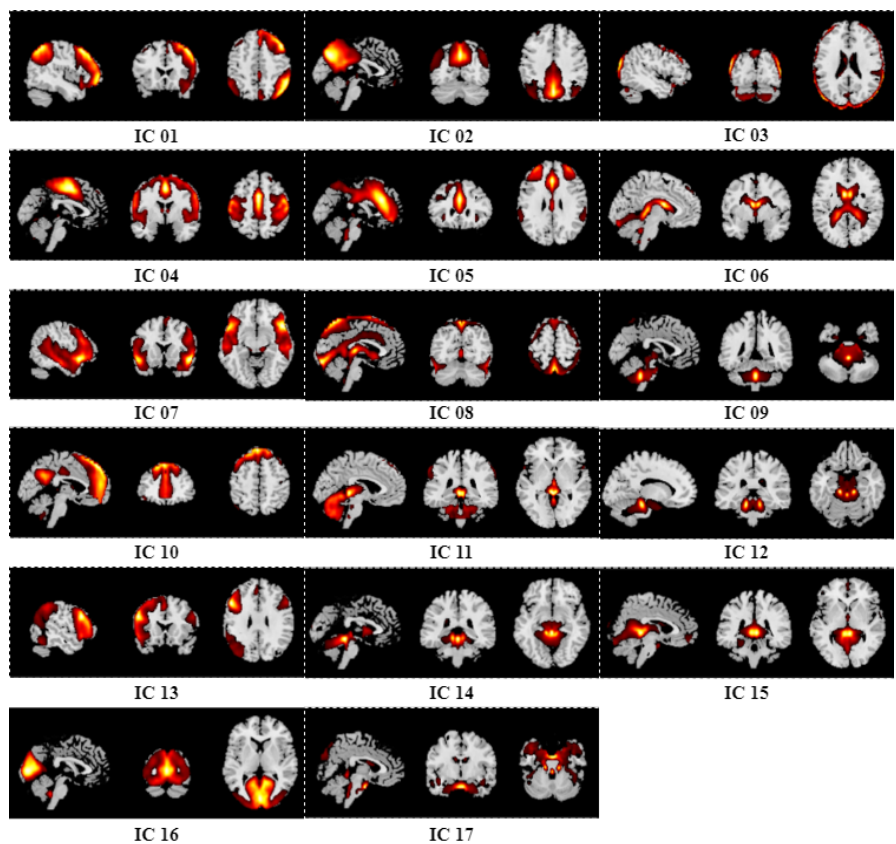

**Figure A2.** Results of spatial maps after group ICA with estimated order number equal to 17.

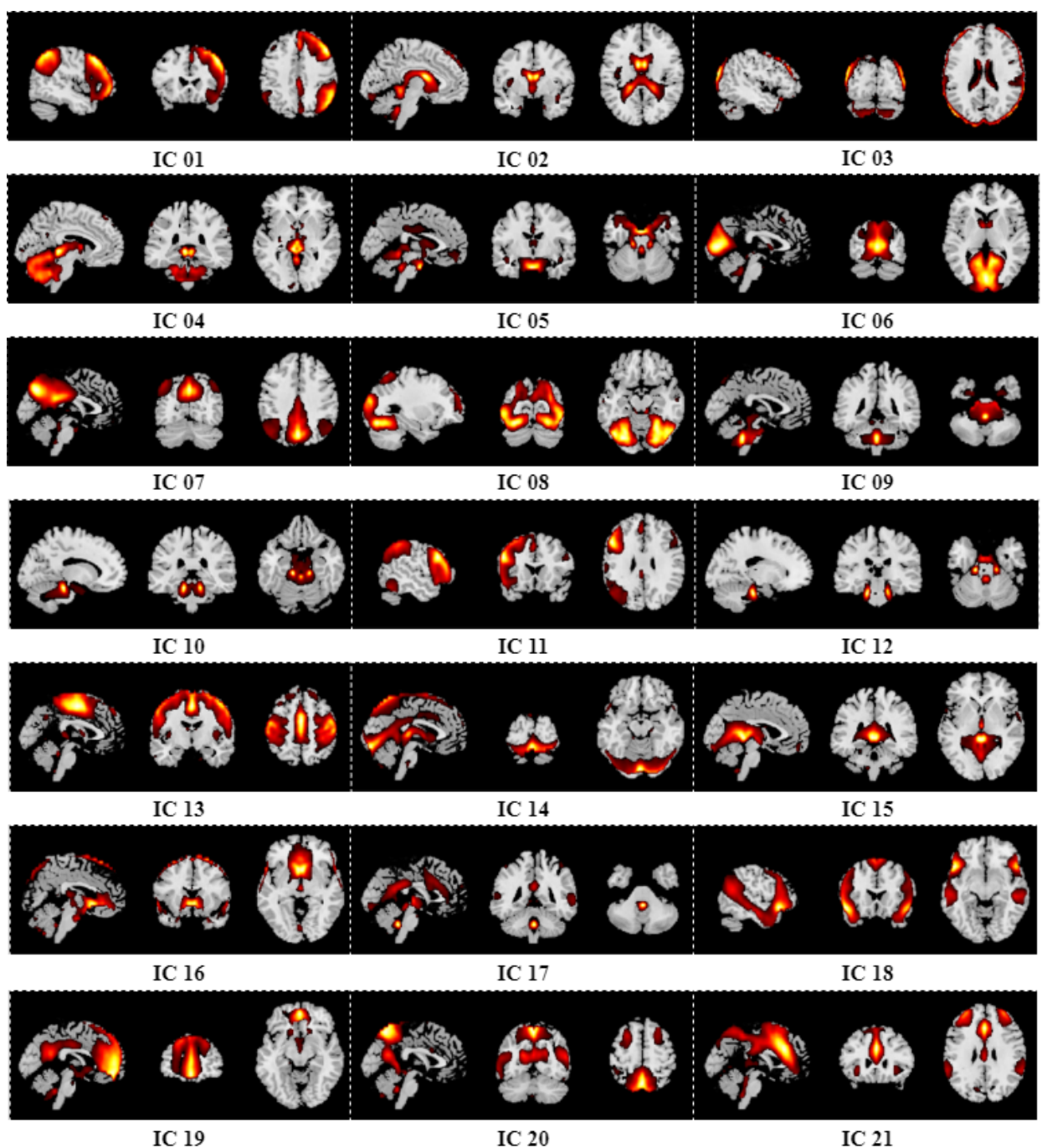

**Figure A3.** Results of spatial maps after group ICA with order number equal to 21.

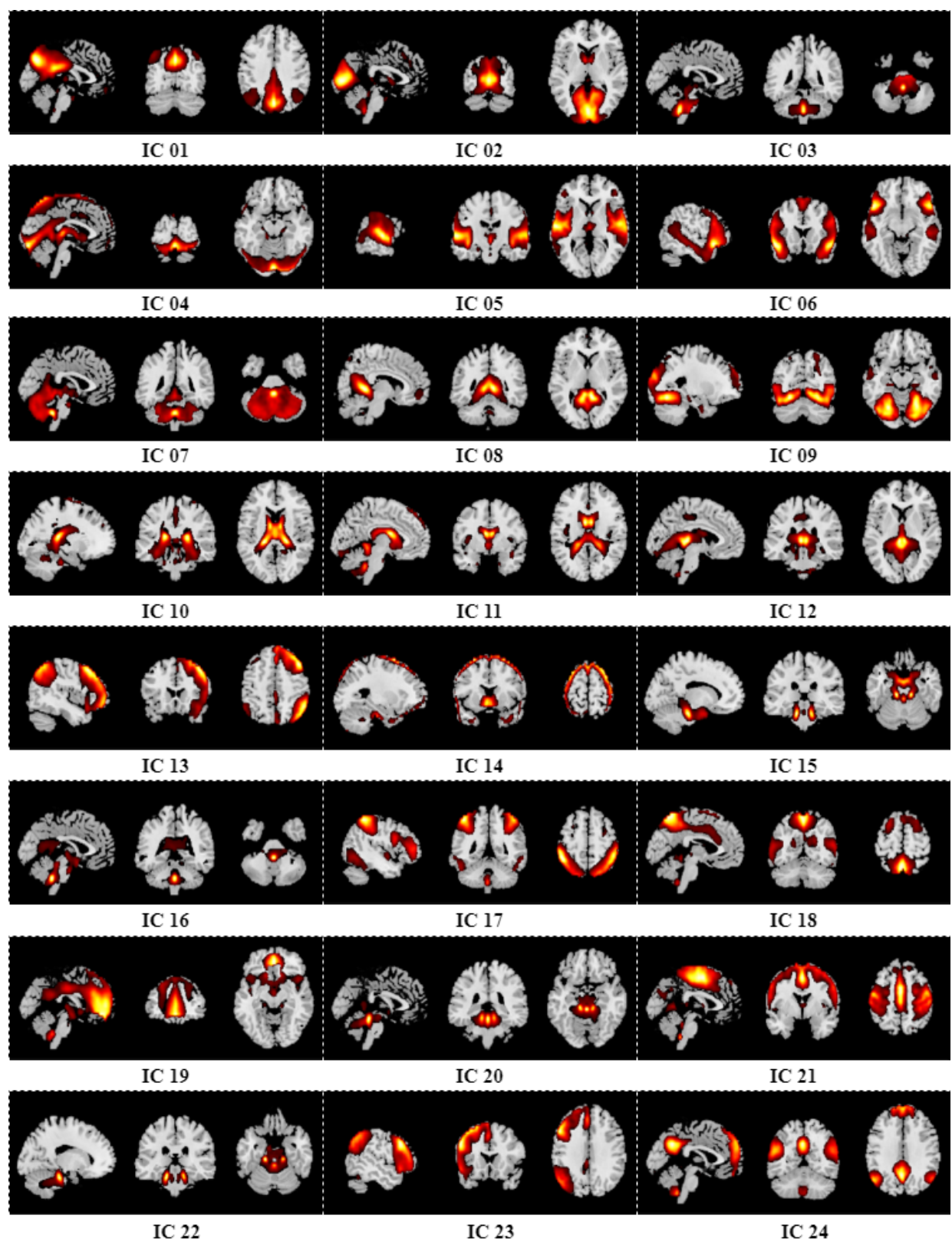

**Figure A4.** Results of spatial maps after group ICA with order number equal to 24.

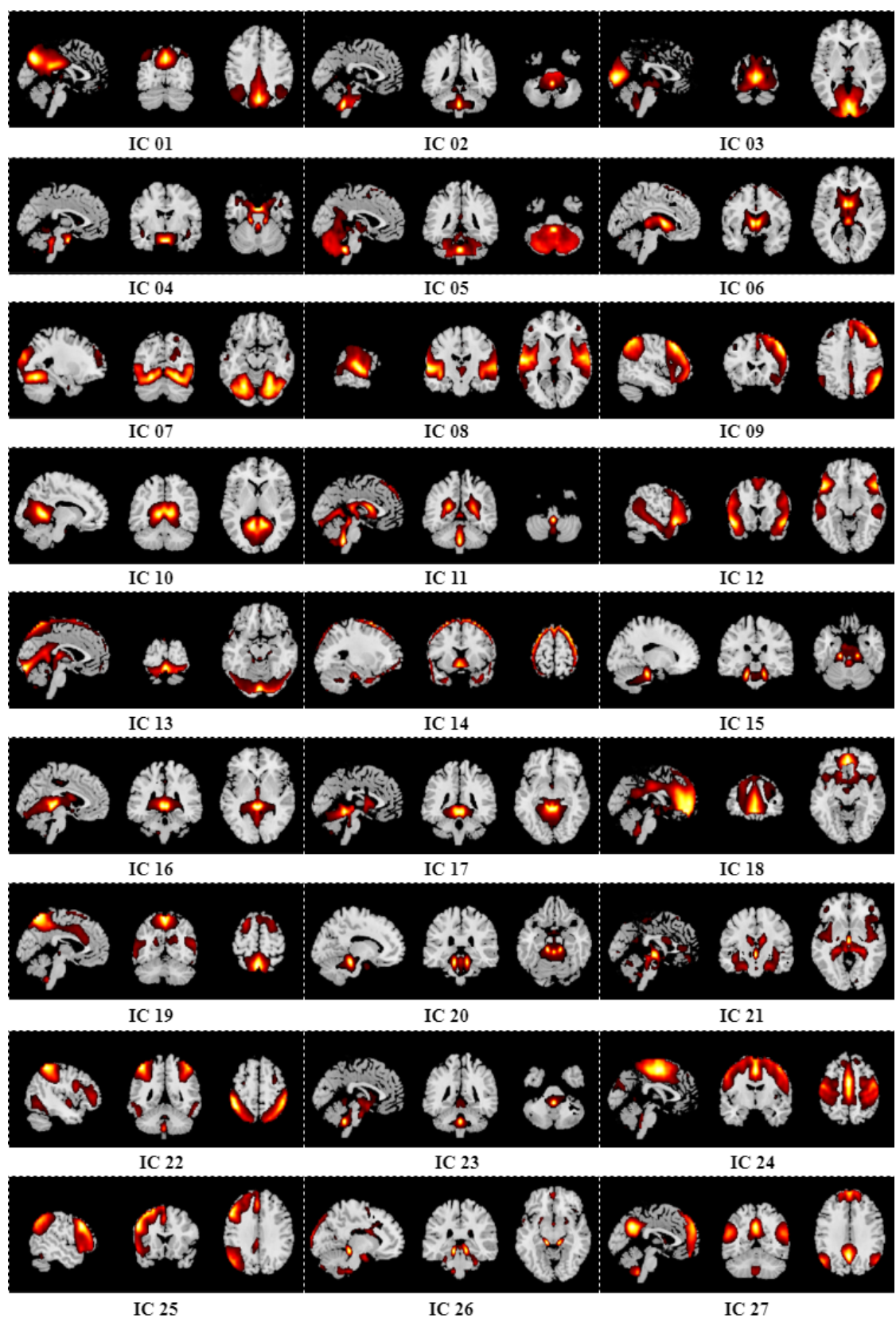

**Figure A5.** Results of spatial maps after group ICA with order number equal to 27.